# Degenerate Rindler Vacuum

M. I. Beciu

Dept. of Physics, Technical University,
B-d Lacul Tei 124, sect. 2
Bucharest, Romania
e-mail: beciu@hidro.utcb.ro

Abstract

. We combine the creation and annihilation operators of a scalar field in Minkowski spacetime to obtain the generators of the SU(1,1). We show that the transformation between Minkowski and Rindler vacua can be represented by (pseudo)rotation in the space of the SU(1,1) group. The root of this fact originates from the coordinate transformation between Minkowski and Rindler manifolds that can also be represented as a (pseudo)rotation in SU(1,1) group. The group structure reveals that an independent (pseudo)rotation in the group space must exist which leads to a new vacuum state, independent Rindler vacuum. The newly defined state shares many of the properties of the Rindler vacuum and, in particular, the theorem of thermalization remains valid for it.

The discovery of a simple relationship of proportionality between temperature and the acceleration of a uniformly accelerated observer, first in the context of black holes [1], then for accelerated observers in Minkowski space-time [2], raised hopes for the unveiling of a deep connection between thermodynamics, quantum mechanics and eventually gravitation. An easy way to see that is by computing the vacuum expectation value of Rindler particle number operator in Minkowski vacuum; the distribution is Planckian (though only in even dimension spacetimes and for massless fields [3]). Even more, with these restrictions, the root mean square fluctuations are thermal.

Here we are concerned with the algebra of the creation and annihilation operators of a scalar field in Minkowski and Rindler vacua. We show these operators can be combined so that they give the generators of the SU(1,1) group. In this case, the transformation between the Minkowski and Rindler vacua is nothing else but a (pseudo)-rotation in the space of the SU(1,1) group. This fact is intimately connected to the relation between the Minkowski and Rindler coordinates which transform under the fundamental representation of the group in the same way. In turn, this allows us to prove that the Rindler vacuum is not unique.

To start with, we recall the connection between the Minkowski and Rindler vacua. They are linked by the transformation

$$\left|O_R\right\rangle = exp(-\sum_k \Phi(\omega)(d_k^{(1)+}d_{-k}^{(2)+} - d_k^{(1)}d_{-k}^{(2)}))\cdot\left|O_M\right\rangle \tag{1}$$

The $d_k^+, d_k$ are the creation and annihilation operators respectively, for the Minkowski vacuum and they are labelled by (1) and (2), corresponding to the two causally disconnected right and left Rindler wedges, respectively. They obey the standard commutation relations and the operators in the two Rindler regions commute. The form of $\Phi(\omega)$ is dictated by the explicit Bogoliubov transformation between the Rindler and Minkowski modes and it turns out to be [4].

$$tanh(\Phi(\omega)) = exp(-\pi\omega / a) \tag{2}$$

with $a$, the constant proper acceleration of an observer in hyperbolic motion. In is worth mentioning that in the thermofield approach [5] a relation similar to (2) (with the argument of the exponential, $-\beta\omega/2$) results from the requirement of Gibbs potential optimisation under constant energy. Relation (2) is instrumental in obtaining the Planckian distribution for the Rindler particle number in Minkowski vacuum.

Now we remark that the transformation (1) can also be written as

$$\left|O_R\right\rangle = exp(-2is_y\cdot\Phi(\omega))\cdot\left|O_M\right\rangle \tag{3}$$

where

$$s_y = -\tfrac{i}{2}\sum_k(d_k^{(1)+}d_{-k}^{(2)+} - d_k^{(1)}d_{-k}^{(2)}) \equiv -\tfrac{i}{2}\sum_k(e_k^+ - e_k) \tag{4}$$

The $d_k^+, d_k$ are the creation and annihilation operators respectively, for the Minkowski vacuum and they are labelled (1) and (2) corresponding to the two causally disconnected right and left Rindler wedges, respectively. The operators $d_k^+, d_k$ obey the standard commutation relations

As can be noticed from (4), $s_y$ involves the ladder operators $e_k^+, e_k$ which can be interpreted as creation and annihilation operators of a pair of particles of definite and opposite momenta, each member of the pair belonging to one of the two Rindler wedges.

Another independent combination of $e_k^+$'s and $e_k$'s is the symmetrized product

$$s_x = \tfrac{1}{2}\sum_k(d_k^{(1)+}d_{-k}^{(2)+} + d_k^{(1)}d_{-k}^{(2)}) \equiv \sum_k s_{xk} \tag{5}$$

Finally, the last independent combination is given by the commutator of $e_k^+$'s, $e_k$'s

$$s_z = \tfrac{1}{2}\sum_k [e_k, e_k^+] = \tfrac{1}{2}\sum_k \ (1 + n_k^{(1)} + n_{-k}^{(2)}) \equiv \sum_k s_{zk} \tag{6}$$

where $n_k = d_k^+ d_k$ denotes the particle number operator. The newly built operators obey, mode by mode, the following commutation relations.

$$[s_{xk}, s_{yl}] = -is_{zk} \cdot \delta_{kl} \ ; \qquad [s_{yk}, s_{zl}] = is_{xk} \cdot \delta_{kl} \ ; \qquad [s_{zk}, s_{xl}] = is_{yk} \cdot \delta_{kl} \tag{7}$$

where no summation is implied on double indices. The presence of the delta symbol indicates simply that the operators $s_{xk}, s_{yk}, s_{zk}$ for different modes are independent. For a given *k*, the above commutation relations represent the algebra of the SU(1,1) group. The operators $s_{xk}, s_{yk}, s_{zk}$ resemble the spin operators introduced by Anderson in studying the superconducting vacuum [6].

Because of the hyperbolic signature in the group invariant,

$$s^2 = s_z^2 - s_x^2 - s_y^2 \tag{8}$$

we will refer to these operators as pseudospin, to exponentials of the form $\exp(i\alpha \cdot s)$ as pseudo-rotations and to the parameters, simply as angles, though, of course, the group is noncompact. Therefore, using relation (1) or (3) one can interpret the Rindler vacuum as the state obtained by a pseudo-rotation of the Minkowki vacuum along the y -axis in the group space, with an angle $2\,\Phi(\omega)$

It is convenient for later calculation and reference to give the two dimensional representation of the pseudospin operators.

$$s_{xk} = \tfrac{1}{2}\begin{pmatrix} 0 & 1 \\ -1 & 0 \end{pmatrix}; \qquad s_{yk} = \tfrac{-1}{2}\begin{pmatrix} 0 & i \\ i & 0 \end{pmatrix}; \qquad s_{zk} = \tfrac{1}{2}\begin{pmatrix} 1 & 0 \\ 0 & -1 \end{pmatrix} \tag{9}$$

To understand more specifically the role played by the group SU(1,1), let us remark first that upon the pseudo rotation (1) the annihilation and creation operators $c_k^{(1)}, c_k^{(1)+}$ and $c_k^{(2)}, c_k^{(2)+}$ for the Rindler vacuum transform as:

$$c_k^{(1)} = \cosh(\Phi) \cdot d_k^{(1)} + \sinh(\Phi) \cdot d_{-k}^{(2)+} \tag{10}$$

and so on, whence the spin operators become under the same pseudo-rotation

$$s_{zk}^R = s_{xk} \cdot \sinh 2\Phi + s_{zk} \cdot \cosh 2\Phi \ ; \ \ s_{xk}^R = s_{xk} \cdot \cosh 2\Phi + s_{zk} \cdot \sinh 2\Phi \ ;$$
$$s_{yk}^R = s_{yk} \tag{11}$$

The pseudospin operators $s_{xk}^R, s_{yk}^R, s_{zk}^R$ are constructed according to the same recipe as their counterparts for Minkowski vacuum (relations (3), (4) and (5)) but from of the operators $c_k^{(1)}, c_k^{(1)+}$ and $c_k^{(2)}, c_k^{(2)+}$. The Minkowki vacuum state is an eigenfunction of the $s_{zk}$ with eigenvalue one half. The same is true for the Rindler vacuum, eigenfunction which is an eigenfunction of $s_{zk}^R$ with eigenvalue one half.

To understand better the origin where the pseudo-rotation stems from we appeal to the transformation relating the Minkowski coordinates *X,T* and the Rindler coordinates $\zeta, \tau$

$$X = \pm\zeta \cosh(a\tau) \quad ; \quad T = \pm\zeta \sinh(a\tau) \tag{12}$$

with the upper signs for the right wedge. It is easy to see that a $\tau$ - translation is equivalent to a Lorentz boost in the *x-t* plane. We will write (12) so that to render evident the SU(1,1) group transformation..

$$u_n = \begin{pmatrix} x \\ t \end{pmatrix} = \begin{pmatrix} cosh(a\tau) & sinh(a\tau) \\ sinh(a\tau) & cosh(a\tau) \end{pmatrix} \cdot \begin{pmatrix} \pm 1 \\ 0 \end{pmatrix} = exp(2ia\tau \cdot s_y) \cdot \begin{pmatrix} \pm 1 \\ 0 \end{pmatrix} \quad (13)$$

where in the last equality we used the representation (9). The spinor $u_n$ is normalised using the scalar product $2u^* s_z u$ appropriate for the SU(1,1) group. Taking also into account that $\zeta^2 = x^2 - t^2$, the coordinate $\zeta$ disappears completely from expression (13). So, in fact the spinor $u_n$ is

$$u_n = \frac{1}{\sqrt{X^2 - T^2}} \begin{pmatrix} X \\ T \end{pmatrix}$$

The quantity $\begin{pmatrix} 1 \\ 0 \end{pmatrix}$ is a unit spinor, taken at the initial moment $t=0$ and pointing up, along $z$-direction in the group space, while $\begin{pmatrix} -1 \\ 0 \end{pmatrix}$ is the same but after a $2\pi$ rotation around $z$. Relation (13) simply says that the coordinates restricted to Rindler manifold can be obtained from these spinors by performing a pseudo-rotation of angle $-2a\tau$ along the O$y$ axis in group space Correspondingly, the ground state of a field confined to Rindler manifold is obtained from Minkowski ground state by a pseudo-rotation along the same axis

For further reference we also write the transformation in the Euclidian sector (Wick rotation in the time complex plane) $T_i=-iT$

$$X = \pm\zeta\, cosh(a\tau)) \quad ; \quad T_i = \mp i\zeta\, sinh(a\tau) \quad (14)$$

The above can be cast again into a form revealing a pseudo-rotation in the SU(1,1) group space.

$$u_n = \frac{1}{\zeta} \begin{pmatrix} X \\ T_i \end{pmatrix} = \begin{pmatrix} cosh(a\tau) & i\, sinh(a\tau) \\ -i\, sinh(a\tau) & cosh(a\tau) \end{pmatrix} \cdot \begin{pmatrix} \pm 1 \\ 0 \end{pmatrix} = exp(2ia\tau \cdot s_x) \cdot \begin{pmatrix} \pm 1 \\ 0 \end{pmatrix} \quad (15)$$

where again we used the representation (9) and the normalization condition.

It would be interesting to compute the pseudospin mean values $s^R_{xk}, s^R_{yk}, s^R_{zk}$ in Minkowski vacuum state. Using (11), it results:

$$\langle O_M | s^R_{zk} | O_M \rangle = \tfrac{1}{2} cosh(2\Phi(\omega)) = \tfrac{1}{2} \cdot cotanh(\pi\omega / a) = \frac{1}{2} + \frac{1}{exp(2\pi\omega / a) - 1} \quad (16a)$$

$$\langle O_M | s^R_{xk} | O_M \rangle = \tfrac{1}{2} \cdot sinh(2\Phi(\omega)) = (2\, sinh(\pi\omega / a))^{-1} \quad (16b)$$

$$\langle O_M | s^R_{yk} | O_M \rangle = 0 \quad (16c)$$

This is one of our main results. Relations (16) support the following interpretation: When the vacuum expectation value of $s^R_{zk}$ is computed in Rindler vacuum its value is one half. When the vacuum state is Minkowski, namely a state is rotated in group space around $Oy$, the value is no longer one half but it is given by (16a) and the state does not point up any longer along $z$ but the $x$ component of the pseudospin acquires a non null value. Relation (16a) displays the celebrated result concerning the Bose distribution of temperature $T=a/(2\pi,)$ including also, the zero point spectrum. It shows that the pseudo-rotation (1) along the $y$-axis converts the zero point oscillations into black body oscillations. At the same time, the $x$ component of the pseudospin develops a nonvanishing vacuum expectation value, relation (16b), which describes the fluctuations in particle number, more precisely the root mean square fluctuations.

At this point it is natural to ask why the pseudo-rotation around the y-axis should be privileged over a pseudo-rotation along the *x*-axis. From the point of view of the signature in relation (8) the *x* and *y* directions are equivalent. Defining a new vacuum state, pseudo-rotated, with respect to the Minkowski state with angle .-2 $\Phi(\omega)$ around the *x* direction (in the group space),

$$\left|O_B\right\rangle = exp(2i\Phi(\omega)\cdot s_x)\left|O_M\right\rangle \tag{17}$$

and following the same steps as in (9), (10), we obtain

$$b_k^{(1)} = cosh(\Phi)\cdot d_k^{(1)} + i\,sinh(\Phi)\cdot d_{-k}^{(2)+} \tag{18}$$

and so on for the annihilation and creation operators of the new vacuum state. The transformation rule, for the pseudospin operators assigned to $\left|O_B\right\rangle$, become

$$s_{zk}^B = s_{yk}\cdot sinh\,2\Phi + s_{zk}\cdot cosh\,2\Phi\,;$$

$$s_{yk}^B = s_{yk}\cdot cosh\,2\Phi + s_{zk}\cdot sinh\,2\Phi\,; \qquad s_{xk}^B = s_{xk} \tag{19}$$

Computing now the Minkowski vacuum expectation values of the pseudospin operators, we meet a surprise.

$$\left\langle O_M\left|s_{zk}^B\right|O_M\right\rangle = \tfrac{1}{2}cosh(2\Phi(\omega)) = \tfrac{1}{2}\cdot cotanh(\pi\omega/a)$$

$$\left\langle O_M\left|s_{yk}^B\right|O_M\right\rangle = \tfrac{1}{2}\cdot sinh(2\Phi(\omega)) = (2\,sinh(\pi\omega/a))^{-1} \tag{20}$$

$$\left\langle O_M\left|s_{xk}^B\right|O_M\right\rangle = 0$$

Again the $s_z$ component develops a Bose-like vacuum expectation value, while the root mean square fluctuations are obtained now for the $s_y$ component. A rather long but straightforward calculation shows that relation (17) can also be rewritten as

$$\left|O_M\right\rangle = \prod_k (cosh(\Phi(\omega)))^{-1}\cdot exp(-i\,tanh(\Phi(\omega)b_k^{(1)+}b_k^{(2)+})\left|O_B\right\rangle \tag{21}$$

Then, an operator $P(b_k^{(1)+}, b_k^{(1)})$ which depends only on the creation and annihilation operators of the Rindler region I, will have an expectation value

$$\left\langle O_M\left|P(b_k^{(1)+}, b_k^{(1)})\right|O_M\right\rangle = \prod_k (1-exp(2\pi\omega/a))\cdot\sum_n exp(-2n\pi\omega/a)\cdot\left\langle n_k^1\left|P\right|n_k^1\right\rangle \tag{22}$$

where we made use of the explicit value of $\Phi(\omega)$ given by (2). In other words, the thermalization theorem [7] holds also true for the $\left|O_B\right\rangle$ vacuum. Mathematically, this result is due to the fact that the pseudospin operators, consisting in pairs of annihilation and creation operators do not sense the difference between an Ox and Oy pseudorotation so that relations (11) and (19) are formally the same.

One can associate the new vacuum state $\left|O_B\right\rangle$ with the Euclidian coordinates (14). The latter are obtained from the unit spinor by a pseudo-rotation around Ox axis in group space, relation (15). Likewise, the state $\left|O_B\right\rangle$ is the result of a pseudo-rotation along the same axis, relation (17).

All the above results hold also true in the black hole case, when one disregards the complications arising from the spherical part of the metric, because the Schwartschild and Kruskal vacua stand in the same relation (1) as the Rindler and Minkowski vacua, respectively.

It has been known for a long time that in flat space there are two ground states for a quantum field, the Minkowski vacuum associated with inertial frames and the Rindler vacuum associated with uniformly accelerated frames. In general, one can pass from inertial to accelerated frames using the conformal group, in particular the acceleration transformation. However, the Rindler transformations (14) are special, they do not form a subgroup of the conformal group and cannot be written as a particular case of the acceleration transformation [8] Instead, the Rindler coordinate transformations can be written, as relation (13) shows, as a SU(1,1) group transformation.

The old result of thermal character of particles in Rindler manifold has been displayed here in form of pseudospin expectation values. We think this picture has certain advantages, at least in terms of intuitivity and visualization.

On the other hand, the rather unexpected result (20) establishes itself as a alternative explanation of why working in Euclidian time one gets the same result as for Lorentzian one.

In quantum field theory, the degeneracy of the ground state is usually the mark of a symmetry breakdown. The existence of the vacuum state $|O_B\rangle$, hence the degeneracy of the ground state for a field confined to the Rindler manifold, signals the occurrence of a broken symmetry.. The study of this problem is in progress.

Thanks are due to G. Ghika and A Sandulescu for critical reading of the manuscript.

## Appendix

For brevity we skipped some calculations. We give them here in some detail.

1. We prove relations (10) and (11) in text

$$c_k^{(1)} = e^{-2i\Phi S_y}\, d_k^{(1)}\, e^{2i\Phi S_y} \tag{1}$$

Using Campbell–Baker formula

$$e^{L} A\, e^{-L} = A + [L, A] + \frac{1}{2}[L,[L,A]]\ldots \tag{2}$$

with $L = -2i\Phi S_y$, we get

$$c_k^{(1)} = d_k^{(1)} - 2i\Phi\left[s_y, d_k^{(1)}\right] + \frac{1}{2}(2i\Phi)^2 [s_y,[s_y,d_k]] + \ldots \tag{3}$$

$$\left[s_y, d_k^{(1)}\right] = -\frac{1}{2}\sum_m \left[d_m^{(1)+} d_{-m}^{(2)+} - d_m^{(1)} d_{-m}^{(2)}, d_k^{(1)}\right] =$$

$$= -\frac{1}{2}\sum_m \left[d_m^{(1)+}, d_k^{(1)}\right]\cdot d_{-m}^{(2)+} = \frac{1}{2}\sum_m \delta_{mk} d_{-m}^{(2)+} = \frac{i}{2}\, d_{-k}^{(2)+} \tag{4}$$

$$\left[s_y,\left[s_y, d_k^{(1)}\right]\right] = -\frac{i^2}{4}\sum_m \left[d_m^{(1)+} d_{-m}^{(2)+} - d_m^{(1)} d_{-m}^{(2)}, d_{-k}^{(1)}\right] =$$

$$= -\frac{1}{4}\sum d_m^{(1)}\left[d_{-m}^{(2)}, d_{-k}^{(2)+}\right] = -\frac{i}{4}\, d_k^{(1)} \tag{5}$$

Replacing (4) and (5) in (3):

$$c_k^{(1)} = d_k^{(1)} + \Phi\, d_{-k}^{(2)+} + \frac{\Phi^2}{2}\, d_k^{(1)} + \ldots = d_k^{(1)} \cosh\Phi + d_{-k}^{(2)+} \sinh\Phi \tag{6}$$

$$c_k^{(1)+} = d_k^{(1)+} \cosh\Phi + d_{-k}^{(2)} \sinh\Phi$$

$$c_k^{(2)} = d_k^{(2)} \cosh\Phi + d_{-k}^{(1)+} \sinh\Phi$$

$$c_k^{(2)+} = d_k^{(2)+} \cosh\Phi + d_{-k}^{(1)} \sinh\Phi$$

We compute now the pseudospin operators

$$s_{zk}^{(R)} = \frac{1}{2}\left(1 + c_k^{(1)+} c_k^{(1)} + c_{-k}^{(2)+} c_{-k}^{(2)}\right)$$

$$2 s_{zk}^{(R)} = \left(1 + \left(d_k^{(1)+} \cosh\Phi + d_{-k}^{(2)} \sinh\Phi\right)\left(d_k^{(1)} \cosh\Phi + d_{-k}^{(2)+} \sinh\Phi\right) +\right.$$

$$\left. + \left(d_{-k}^{(2)+} \cosh\Phi + d_k^{(1)} \sinh\Phi\right)\left(d_{-k}^{(2)} \cosh\Phi + d_k^{(1)+} \sinh\Phi\right)\right) =$$

$$= 1 + d_k^{(1)+} d_k^{(1)} \cosh^2\Phi + d_k^{(1)+} d_{-k}^{(2)+} \sinh\Phi \cosh\Phi + d_{-k}^{(2)} d_k^{(1)} \sinh\Phi \cosh\Phi +$$

$$+ d_{-k}^{(2)} d_{-k}^{(2)+} \sinh^2\Phi + d_{-k}^{(2)+} d_{-k}^{(2)} \cosh^2\Phi + d_{-k}^{(2)+} d_k^{(1)+} \sinh\Phi \cosh\Phi +$$

$$+d_k^{(1)}d_{-k}^{(2)}\sinh\Phi\cosh\Phi+d_k^{(1)}d_k^{(1)+}\sinh^2\Phi \quad (7)$$

Using the commutation relation

$$\left[d_k, d_k^+\right]=1 \quad (8)$$

and the definitions of pseudospin operators, rels. (4) and (5), we get

$$2s_{zk}^{(R)}=1+d_k^{(1)+}d_k^{(1)}\left(\cosh^2\Phi+\sinh^2\Phi\right)+\sinh^2\Phi+$$

$$+2\left(d_{-k}^{(2)}d_k^{(1)}+d_k^{(1)+}d_{-k}^{(2)+}\right)\sinh\Phi\cosh\Phi+d_{-k}^{(2)+}d_{-k}^{(2)}\left(\cosh^2\Phi+\sinh^2\Phi\right)+\sinh^2\Phi=$$

$$=\cosh 2\Phi\left(1+d_k^{(1)+}d_k^{(1)}+d_{-k}^{(2)+}d_{-k}^{(2)}\right)+\sinh 2\Phi\left(d_{-k}^{(2)}d_k^{(1)}+d_k^{(1)+}d_{-k}^{(2)+}\right)=$$

$$=2s_{zk}\cosh 2\Phi+2s_{xk}\sinh 2\Phi \quad (9)$$

$$2s_{xk}^{(R)}=\left(c_k^{(1)+}c_{-k}^{(2)+}+c_k^{(1)}c_{-k}^{(2)}\right)=$$

$$=\left(d_k^{(1)+}\cosh\Phi+d_{-k}^{(2)}\sinh\Phi\right)\cdot\left(d_{-k}^{(2)+}\cosh\Phi+d_k^{(1)}\sinh\Phi\right)+$$

$$+\left(d_k^{(1)}\cosh\Phi+d_{-k}^{(2)+}\sinh\Phi\right)\cdot\left(d_{-k}^{(2)}\cosh\Phi+d_k^{(1)+}\sinh\Phi\right)=$$

$$=2\left(d_k^{(1)+}d_{-k}^{(2)+}\cosh^2\Phi+d_{-k}^{(2)}d_k^{(1)}\sinh^2\Phi\right)+d_k^{(1)+}d_k^{(1)}\cosh\Phi\sinh\Phi+$$

$$+d_{-k}^{(2)}d_{-k}^{(2)+}\sinh\Phi\cosh\Phi=$$

$$=2\left(s_{zk}\sinh 2\Phi+s_{xk}\cosh 2\Phi\right) \quad (10)$$

Along the same lines one proves relations (18), (19) in text.

$$b_k^{(1)}=e^{2i\Phi S_x}\,d_k^{(1)}\,e^{-2i\Phi S_x} \quad ; \quad L=-2i\Phi S_x$$

$$b_k^{(1)}=d_k^{(1)}+2i\Phi\left[s_{xk},d_k^{(1)}\right]+\frac{1}{2}(2i\Phi)^2\left[s_{xk},\left[s_{xk},d_k^{(1)}\right]\right]+\ldots$$

$$\left[s_{xk},d_k^{(1)}\right]=-\frac{1}{2}\,d_{-k}^{(2)+}$$

$$\left[s_{xk},\left[s_{xk},d_k^{(1)}\right]\right]=-\frac{1}{4}\,d_k^{(1)}$$

$$b_k^{(1)}=d_k^{(1)}-i\Phi\,d_{-k}^{(2)+}+\frac{\Phi^2}{2}\,d_k^{(1)}-\frac{i}{3!}\,\Phi^3\,d_{-k}^{(2)+}+\ldots=$$

$$=\cosh\Phi\,d_k^{(1)}-i\sinh\Phi\,d_{-k}^{(2)+}$$

etc.

2. We derive relations (13) and (15)

$$\exp(2ia\tau s_y)=\begin{pmatrix}\cosh a\tau & \sinh a\tau\\ \sinh a\tau & \cosh a\tau\end{pmatrix} \quad (1)$$

and

$$exp(2ia\tau s_x)=\begin{pmatrix} cosh\,a\tau & i\,sinh\,a\tau \\ -i\,sinh\,a\tau & cosh\,a\tau \end{pmatrix} \tag{2}$$

We utilize the fundamental representations (9)

$$s_x=\frac{1}{2}\begin{pmatrix} 0 & 1 \\ -1 & 0 \end{pmatrix} \quad ; \quad s_y=-\frac{1}{2}\begin{pmatrix} 0 & i \\ i & 0 \end{pmatrix} \tag{3}$$

$$exp\left(ia\tau\begin{pmatrix} 0 & 1 \\ -1 & 0 \end{pmatrix}\right)=\sum_k \frac{(ia\tau)^k}{k!}\begin{pmatrix} 0 & 1 \\ -1 & 0 \end{pmatrix}^k = \tag{2'}$$

$$=\sum_{2k}\frac{(ia\tau)^{2k}}{k!}\begin{pmatrix} 0 & 1 \\ -1 & 0 \end{pmatrix}^{2k}+\sum_{2k+1}\frac{(ia\tau)^{2k+1}}{k!}\begin{pmatrix} 0 & 1 \\ -1 & 0 \end{pmatrix}^{2k+1}$$

$$\begin{pmatrix} 0 & 1 \\ -1 & 0 \end{pmatrix}^{2k}=-\begin{pmatrix} 0 & 1 \\ 1 & 0 \end{pmatrix}^{k} \quad ; \quad \begin{pmatrix} 0 & 1 \\ -1 & 0 \end{pmatrix}^{2k+1}=-\begin{pmatrix} 0 & 1 \\ -1 & 0 \end{pmatrix}$$

$$exp\left(ia\tau\begin{pmatrix} 0 & 1 \\ -1 & 0 \end{pmatrix}\right)=-\sum_{2k}\frac{(ia\tau)^{2k}}{k!}\begin{pmatrix} 0 & 1 \\ 1 & 0 \end{pmatrix}^{k}+\sum_{2k+1}\frac{(ia\tau)^{2k+1}}{k!}\begin{pmatrix} 0 & 1 \\ -1 & 0 \end{pmatrix}(-1)^k = \tag{2''}$$

$$=\begin{pmatrix} cosh\,a\tau & 0 \\ 0 & cosh\,a\tau \end{pmatrix}+i\begin{pmatrix} 0 & sinh\,a\tau \\ -sinh\,a\tau & 0 \end{pmatrix}$$

$$exp\left(-ia\tau\begin{pmatrix} 0 & i \\ i & 0 \end{pmatrix}\right)=exp\left(a\tau\begin{pmatrix} 0 & 1 \\ 1 & 0 \end{pmatrix}\right)=\sum_k \frac{(a\tau)^k}{k!}\begin{pmatrix} 0 & 1 \\ 1 & 0 \end{pmatrix}^k \tag{1'}$$

$$\begin{pmatrix} 0 & 1 \\ 1 & 0 \end{pmatrix}^{2k}=I \quad ; \quad \begin{pmatrix} 0 & 1 \\ 1 & 0 \end{pmatrix}^{2k+1}=\begin{pmatrix} 0 & 1 \\ 1 & 0 \end{pmatrix}$$

$$exp\left(-ia\tau\begin{pmatrix} 0 & i \\ i & 0 \end{pmatrix}\right)=\sum_{2k}\frac{(a\tau)^{2k}}{(2k)!}\begin{pmatrix} 1 & 0 \\ 0 & 1 \end{pmatrix}+\sum_{2k+1}\frac{(a\tau)^{2k+1}}{(2k+1)!}\begin{pmatrix} 0 & 1 \\ 1 & 0 \end{pmatrix}=$$

$$=\begin{pmatrix} cosh\,a\tau & 0 \\ 0 & cosh\,a\tau \end{pmatrix}+\begin{pmatrix} 0 & sinh\,a\tau \\ sinh\,a\tau & 0 \end{pmatrix}$$

On the other hand, the norm of $\begin{pmatrix} X \\ T \end{pmatrix}\frac{1}{\zeta}$ is

$$\begin{pmatrix} cosh\,a\tau & sinh\,a\tau \end{pmatrix}\begin{pmatrix} 1 & 0 \\ 0 & -1 \end{pmatrix}\begin{pmatrix} cosh\,a\tau \\ sinh\,a\tau \end{pmatrix}=1$$

The norm of $\begin{pmatrix} X \\ T_i \end{pmatrix}\frac{1}{\zeta}$ is

$$\begin{pmatrix} cosh\,a\tau & -i\,sinh\,a\tau \end{pmatrix}\begin{pmatrix} 1 & 0 \\ 0 & -1 \end{pmatrix}\begin{pmatrix} cosh\,a\tau \\ i\,sinh\,a\tau \end{pmatrix}=1$$

3. To prove formula (22), we first simplify the notations

$$r = tanh\,\Phi(\omega) = e^{-\frac{\pi\omega}{a}} \quad ; \quad \frac{1}{cosh^2\,\Phi} = 1 - r^2$$

and we remark that

$$exp\left(-ir\,b^{(1)+}b^{(2)+}\right)\left|O_B\right\rangle = \sum_n (-ir)^n \left|n^{(1)},n^{(2)}\right\rangle_B$$

where

$$\left|n^{(1)},m^{(2)}\right\rangle_B = \frac{\left(b^{(1)+}\right)^n \left(b^{(2)+}\right)^m}{\sqrt{n!}\sqrt{m!}}\left|O_B\right\rangle$$

then

$$\left\langle O_M\right|P\left(b^{(1)+}b^{(1)}\right)\left|O_M\right\rangle = \frac{1}{cosh^2\,\Phi}\cdot\sum_{n,m}(-ir)^n(+ir)^m\left\langle m^{(1)},m^{(2)}\right|\;P\left(b^{(1)+},b^{(1)}\right)\left|n^{(1)},n^{(2)}\right\rangle$$

Because of

$$\sum_m \left\langle m^{(2)}\right|\left|n^{(2)}\right\rangle = \delta_{mn}$$

we get

$$(1-r^2)\sum_n r^{2n}\left\langle n^{(1)}\right|P\left(b^{(1)+},b^{(1)}\right)\left|n^{(1)}\right\rangle =$$

$$= [1 - exp(-2\pi\omega/a)]\sum_n exp(-2\pi n\omega)\left\langle n^{(1)}\right|P\left(b^{(1)+},b^{(1)}\right)\left|n^{(1)}\right\rangle$$

degenerated